\documentclass[12pt]{iopart}
\usepackage{graphicx}

\begin{document}

\title[On the limited amplitude resolution of multipixel Geiger-mode APDs]
{On the limited amplitude resolution of multipixel Geiger-mode APDs}

\author{A~Stoykov$^{1,3}$, Y~Musienko$^{2,4}$, A~Kuznetsov$^3$,
S~Reucroft$^2$ and J~Swain$^2$}

\address{$^1$~Paul Scherrer Institut, CH-5232 Villigen PSI, Switzerland}
\address{$^2$~Department of Physics, Northeastern University,
Boston, MA 02115, USA}
\address{$^3$~Joint Institute for Nuclear Research, 141980 Dubna, Russia}
\address{$^4$~On leave from INR, Moscow}

\ead{alexey.stoykov@psi.ch (A~Stoykov)}

\begin{abstract}
The limited number of active pixels in a Geiger-mode Avalanche Photodiode
(G-APD) results not only in a non-linearity but also in
an additional fluctuation of its response.
Both these effects are taken into account to calculate the amplitude resolution
of an ideal G-APD, which is shown to be finite.
As one of the consequences, the energy resolution of a scintillation detector based
on a G-APD is shown to be limited to some minimum value defined by the number of
pixels in the G-APD.
\end{abstract}


\vspace{2pc}


\maketitle

Multipixel Geiger-mode avalanche photodiodes (G-APDs) \cite{Renker_Beaune2005}
are solid state photodetectors with a fast growing field of applications.
A G-APD can be thought of as a matrix of independent photon counters
(pixels\,/\,cells) connected to a common load.
A photoelectron produced in the G-APD volume by a photon can trigger an
avalanche breakdown in one of the cells (fire the cell).
To an input signal of $n$ photoelectrons
the G-APD responds with an output signal of $N$ fired cells,
the dependence $N = f(n)$ saturates
due to the limited number $m$ of the cells in the device, and the fact that
more than one photon can enter one cell without changing the response of that cell.
The non-linearity of a G-APD response is an established experimental fact
(see, e.g \cite{Dolgoshein_ICFA}).
A correlation between saturation of the G-APD signal and an increase of
its dispersion was also observed \cite{Dolgoshein_ICFA}.

In this work we calculate the amplitude resolution of an ideal G-APD
(which we call the "intrinsic" resolution)
and show that it is limited (non zero) due to an additional
statistical noise in the process of distributing $n$ photoelectrons
over $m$ cells. By an ideal G-APD we understand a device with the excess
noise factor \cite{Musienko_Beaune2005} equal to unity. In particular,
we assume that there is no crosstalk between cells (so it is never the
case that one photon triggers more than one cell) and that each photon incident
hits some cell.

The considered problem is equivalent to a well-known problem in
mathematical statistics of distributing (randomly) $n$ balls (photoelectrons)
into $m$ urns (cells), see e.g. \cite{Urn_Models}.
The number $N$ of urns containing one or more balls is a random variable,
its expected value and variance are:

\begin{equation}
\begin{array}{rcl}
{\bar N} & = & m\,\left[1-(1 - m^{-1})^n\right] \\[2ex]
\sigma_N^2 & = & m\,(m-1)\,(1-2\,m^{-1})^n + m\,(1-m^{-1})^n - m^2\,(1-m^{-1})^{2n}
\end{array}
\label{Eq:n}
\vspace{1ex}
\end{equation}

\noindent
The distribution of $N$ is approximately normal
when $m, n \rightarrow \infty$ and the ratio $\alpha = n/m$ is bounded
\cite{Urn_Models}:

\begin{equation}
\begin{array}{rcl}
{\bar N} & = & m\,(1 - \e^{-\alpha}) \\[2ex]
\sigma_N^2 & = & m\,\e^{-\alpha}\,[1-(1+\alpha)\,\e^{-\alpha}] \ .
\end{array}
\label{Eq:n_inf}
\vspace{1ex}
\end{equation}

In practice the number of cells in a G-APD is usually greater than $\sim 100$,
which justifies using the asymptotic formulae (\ref{Eq:n_inf}) in the following
analysis.

The amplitude resolution of a detector is a measure of its ability to distinguish
between input signals having different amplitudes \cite{Leo}.
For a device with a linear response the resolution can be shown to be:
$R = 2.35 \cdot \sigma/{\bar A}$, where $A$ is the amplitude of the detector
response, and $\bar A$ and $\sigma$ are the mean value and standard deviation
characterizing the distribution of $A$ (which is assumed to be normal).
In case of a non-linear response, the formula has to be modified.
The scheme of our consideration is the following:

\vspace{-1ex}
$$
A_1 \ \stackrel{f}\longrightarrow \ A:({\bar A}, \sigma) \
\stackrel{f^{-1}}\longrightarrow \ A_2:({\bar A_2}, \sigma_2) \ .
$$
To an input signal $A_1$ (constant value) the detector responds with
a signal of amplitude $A$. Due to fluctuations induced by the detector,
$A$ is statistically distributed. If the function $f$
relating $\bar A$ to $A_1$ is known, one can reconstruct an estimate of the input signal
from the mean detector response. The restored amplitude $A_2$ is no longer a constant,
like $A_1$, but is also a random value, with mean and standard deviations
(at $\sigma \ll {\bar A}$):

\begin{equation}
{\bar A}_2 = f^{-1}\,({\bar A}) \equiv A_1 \ , \qquad\qquad
\sigma_2 = \left(f^{-1}\right)^{'}_{\bar A} \cdot \sigma \ .
\label{Eq:A2s2}
\end{equation}

The parameters of the restored distribution, rather than the measured one,
define the amplitude resolution of a detector:

\vspace{-1ex}
\begin{equation}
R = 2.35 \cdot (\sigma_2\,/\,{\bar A}_2) \ .
\label{Eq:R_def}
\vspace{1ex}
\end{equation}

The theoretical function ${\bar N} = f(n)$ is known (\ref{Eq:n_inf}),
and this allows us, using relations (\ref{Eq:A2s2}),
to reconstruct the signal at the G-APD input
($A_2$ is the reconstructed amplitude)
from the measured distribution of $N$
(in (\ref{Eq:A2s2}) ${\bar A} \equiv {\bar N}$, $\sigma \equiv \sigma_N$):

\vspace{-0.5ex}
$$
{\bar A_2} = n \ , \qquad
\sigma_2 = \sigma_N \,\e^\alpha \ .
$$

By substituting ${\bar A_2}$ and $\sigma_2$ in ({\ref{Eq:R_def})
for the intrinsic amplitude resolution $R_0$ of a G-APD we get:

\begin{equation}
\begin{array}{ccl}
R_0 & = & 2.35 \cdot m^{-1/2} \cdot \Phi_0(\alpha) \ ,\\[2ex]
\Phi_0(\alpha) & = & \alpha^{-1}\,(\e^\alpha - 1 - \alpha)^{1/2} \ .
\end{array}
\label{Eq:R0}
\vspace{1ex}
\end{equation}

The function $\Phi_0(\alpha)$ is plotted in Figure~1.
For any given value of $m$ the value of $R_0$ increases
(the resolution becomes worse)
with the ratio $n/m = \alpha$ increasing.

\begin{figure}[htb]
\begin{center}
\includegraphics*[width=8.5cm]{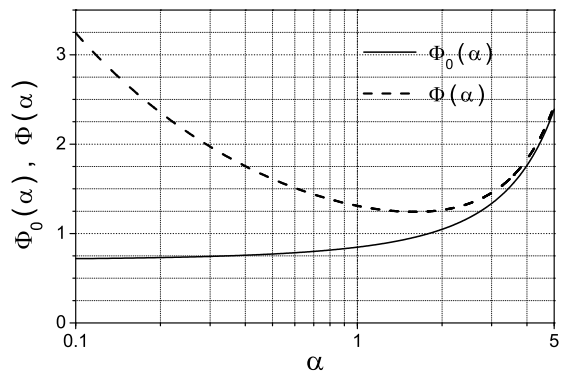}\\
\caption{The functions $\Phi_0(\alpha)$ and $\Phi(\alpha)$ defined in
(\ref{Eq:R0}) and (\ref{Eq:R}).
}
\end{center}
\end{figure}

When a G-APD is used in a scintillation detector the intrinsic resolution of
the G-APD should be summed quadratically
(this we prove in Monte-Carlo simulations below) with the resolution determined
by the statistics of photoelectrons $R_{\rm stat} = 1/\sqrt{\bar n}$
(the mean number of photoelectrons ${\bar n}$ relates to the number
of photons as ${\bar n} = PDF \cdot {\bar n}_{\rm ph}$, where $PDE$ is the photon
detection efficiency \cite{Musienko_Beaune2005}).
Note, that here we assume that the energy resolution of the scintillator
\cite{Dorenbos95} is equal to zero.
The energy resolution of such an ideal scintillation detector is then:

\begin{equation}
\begin{array}{ccl}
R & = & 2.35 \cdot m^{-1/2} \cdot \Phi(\alpha)\ ,\\[2ex]
\Phi(\alpha) & = & \left(\,\alpha^{-1} + \Phi_0^2(\alpha)\,\right)^{1/2} \ .
\end{array}
\label{Eq:R}
\vspace{2ex}
\end{equation}

The function $\Phi(\alpha)$ (see Figure~1)
has minimum value of $\Phi_{\rm min} = 1.2426$ at $\alpha = 1.5936$.
Accordingly, the minimum value for the energy resolution of an ideal G-APD based
scintillation detector (taking into account only the statistics of photoelectrons)
is $R_{\rm min} = 2.92/\sqrt{m}$.

\begin{figure}[htb]
\begin{center}
\includegraphics*[width=10cm]{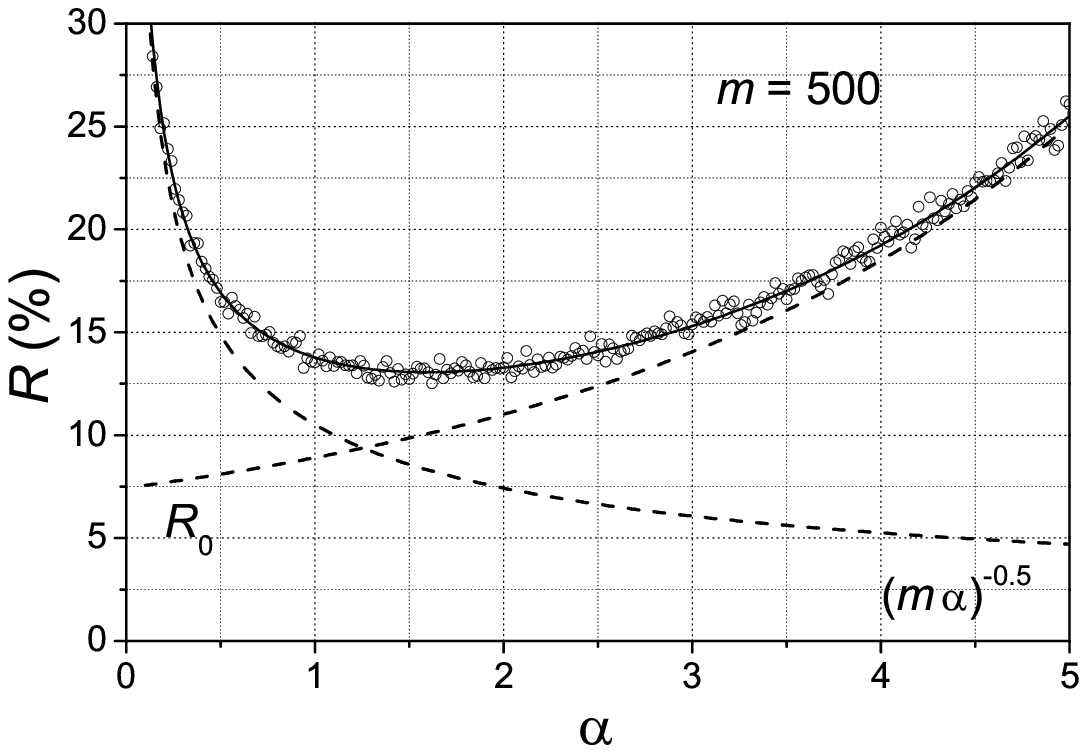}
\includegraphics*[width=10cm]{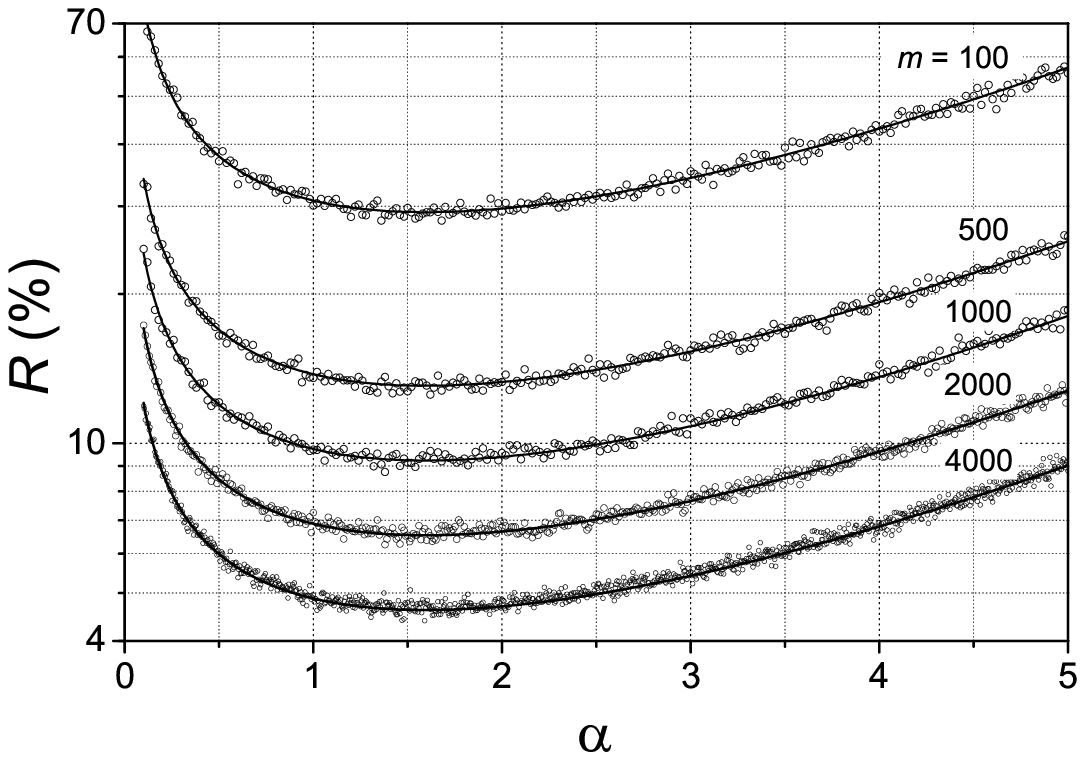}
\caption{
(Top)~Simulated values (dots) of the energy resolution of
an ideal G-APD based scintillation detector as compared with the values (solid line)
calculated according with (\ref{Eq:R}). The dashed lines show two contributions
to the energy resolution: the statistical term ($1/\sqrt{m \, \alpha}$)
and the intrinsic resolution of the G-APD ($R_0$).
(Bottom)~the same for G-APDs with different number of cells $m$.
}
\end{center}
\end{figure}

The obtained result on the energy resolution was verified
in Monte-Carlo simulations. In the simulations the energy resolution of
an ideal scintillation detector was calculated as the follows:
a)~a number of photoelectrons $n$ was randomly generated according to
Poisson statistics;
b)~those $n$ photoelectrons were randomly (uniformly) distributed over
$m$ cells resulting in $N$ occupied (fired) cells.
c)~from the distribution of $N$ the number of photoelectrons (the energy)
and its standard deviation were reconstructed using (\ref{Eq:A2s2}),
and the energy resolution calculated according to (\ref{Eq:R_def}).
As is seen in Figure~2, the energy resolution obtained in simulations is in
agreement with the values calculated according with (\ref{Eq:R}).

Physically, while a larger mean number of photons implies a smaller fluctuation
in that mean, at some point saturation effects dominate (multiple photons
enter single cells) and resolution degrades. This means that some care should
be taken in each application to optimize design taking this into account.

\subsection*{Summary}
The variation of a G-APD response induced by the statistics of distributing
$n$ photoelectrons over $m$ G-APD cells causes its
amplitude resolution to degrade as the ratio $n/m$ increases.
This implies, for example, in that the energy resolution
of a scintillation detector using a G-APD is limited to
$R_{\rm min} = 2.92\,/\,\sqrt{m}$.

\end{document}